\documentclass[twocolumn,superscriptaddress,aps,prb,UTF8]{revtex4-2}
\usepackage{amsmath}
\usepackage{amssymb}
\usepackage{graphicx}
\usepackage{dcolumn}
\usepackage{color}
\usepackage{ulem}
\usepackage{bm}
\usepackage{braket}
\usepackage[colorlinks,citecolor=blue]{hyperref}
\usepackage{multirow}
\usepackage{makecell}
\usepackage{booktabs}
\usepackage{graphicx} 
\usepackage{physics}

\begin{document}

\title{Probing the Topology of Fermionic Gaussian Mixed States with $U(1)$ symmetry by Full Counting Statistics}
\author{Liang Mao}
\affiliation{Institute for Advanced Study, Tsinghua University, Beijing,100084, China}
\author{Hui Zhai}
\email{corresponding author:hzhai@tsinghua.edu.cn}
\affiliation{Institute for Advanced Study, Tsinghua University, Beijing,100084, China}
\affiliation{Hefei National Laboratory, Hefei 230088, China}
\author{Fan Yang}
\email{corresponding author:101013867@seu.edu.cn}
\affiliation{School of Physics, Southeast University, No.2 SEU Road, Nanjing 211189, China}

\date{\today}

\begin{abstract}

Topological band theory has been studied for free fermions for decades, and one of the most profound physical results is the bulk-boundary correspondence. Recently a focus in topological physics is extending topological classification to mixed states. Here, we focus on Gaussian mixed states where the modular Hamiltonians of the density matrix are quadratic free fermion models with $U(1)$ symmetry and can be classified by topological invariants. The bulk-boundary correspondence is then manifested as stable gapless modes of the modular Hamiltonian and degenerate spectrum of the density matrix. In this article, we show that these gapless modes can be detected by the full counting statistics, mathematically described by a function introduced as $F(\theta)$. A divergent derivative at $\theta=\pi$ can be used to probe the gapless modes in the modular Hamiltonian. Based on this, a topological indicator, whose quantization to unity senses topologically nontrivial mixed states, is introduced. We present the physical intuition of these results and also demonstrate these results with concrete models in both one- and two-dimensions. Our results pave the way for revealing the physical significance of topology in mixed states. 

\end{abstract}

\maketitle

{\onecolumngrid
}

\hfill \break

\twocolumngrid
Topology has been extensively used as a tool to characterize the properties of physical systems for decades \cite{Witten16, Wen17,Moessner21,Bernevig,TI1,TI2,TI3}. The topological properties of quantum phases are not just mathematical concepts but also have profound physical consequences. The most well-known physical consequence is the bulk-boundary correspondence, which states that the nontrivial topological invariant of the insulating bulk can ensure a gapless state stable against perturbations at the boundary \cite{Bernevig,TI1,TI2,TI3}. In two dimensions, such a one-dimensional gapless edge state can manifest as quantized conductance in transport experiments, such as quantum anomalous Hall \cite{QAH1,QAH2} and quantum spin Hall effect \cite{QSH1,QSH2,QSH3}. 

Previously, most studies of the topological phases of matter focused on pure states, most relevant to a gapped insulator at equilibrium and with temperatures much lower than the band gap \cite{Witten16, Wen17,Moessner21,Bernevig,TI1,TI2,TI3,QAH1,QAH2,QSH1,QSH2,QSH3}. This is also the situation where most condensed matter experiments on topological materials have been carried out. Nevertheless, there are other situations where the quantum states are intrinsically mixed. These situations include systems described by a finite temperature density matrix  when the temperature is comparable to the band gap, or systems inevitably driven to a mixed state by coupling to an environment. These situations are increasingly important especially for topological phases in platforms such as ultracold atoms \cite{cold_atom_TI}. 

This progress demands extending the studies of topological properties to mixed states, which is a focus of current research in topological physics \cite{top_dens1,top_dens2,top_dens3,top_dens4,top_dens5,top_dens6,top_dens7,top_dens8,top_mod1,top_mod2,top_mod3,top_mod4,top_mod5,top_mod6,top_mod7, ZMHuang24,Huang24,top_lind1,top_lind2,top_lind3,top_dis1,top_dis2,top_dis3,top_dis4,top_dis5,top_dis6,top_dis7,top_dis8,top_dis9,Yin23,Bao23,Fan23,aspt1,aspt2,cenke1,cenke2,interaction1,interaction2,interaction3,interaction4,interaction5}. A mixed state is described by a density matrix $\hat{\rho}$, and a density matrix can be equivalently characterized by its modular Hamiltonian $\hat{G}$ defined as $\hat{\rho}=e^{-\hat{G}}/Z$, where $Z$ is a normalization factor. For a finite temperature thermal state, the modular Hamiltonian is equivalent to the physical Hamiltonian, up to a factor of inverse temperature. While for a general mixed state, these two Hamiltonians are usually different. Since the way to characterize the topology of a quadratic free fermion Hamiltonian is well established \cite{Witten16, Wen17,Moessner21,Bernevig,TI1,TI2,TI3}, it is straightforward to borrow these well established results to study the topology of Gaussian mixed states, whose modular Hamiltonians are quadratic free fermion ones. A topologically nontrivial Gaussian mixed state is characterized by nontrivial topology of its modular Hamiltonian. Such an idea has been explored in many recent papers \cite{top_mod1,top_mod2,top_mod3,top_mod4,top_mod5,top_mod6,top_mod7,ZMHuang24,Huang24}.

An immediate follow-up question is the physical consequence of the nontrivial density matrix topology defined by its modular Hamiltonian. Especially, \textit{whether there exist quantized topological indicators associated with topologically nontrivial mixed states.} We note that there is existing literature discussing the topology of one- and two-dimensional free fermions at finite temperatures, where  quantized observables based on the Zak phase were proposed \cite{top_mod1,top_mod2,top_mod3,top_mod4}. The exact forms of these observables depend on the dimension and the symmetry class of the modular Hamiltonian. In this paper, we propose universal quantized observables that apply to all fermionic Gaussian sates with $U(1)$ symmetry in any dimensions. These observables essentially probe the zero-energy edge modes. Since accidental zero-energy modes not protected by topology are easily removed by perturbations, which are ubiquitous in real systems, these observables can be used as diagnoses of mixed state topology. 

Following the well-established bulk-boundary correspondence, a topologically nontrivial bulk Hamiltonian in $d$-dimensions ensures stable gapless zero-energy modes in its $(d-1)$-dimensional boundary. The bulk-boundary correspondence can be viewed as a mathematical property of any quadratic Hamiltonian, which holds regardless of physical Hamiltonian or modular Hamiltonian. A zero-energy mode in physical Hamiltonian means that adding a particle at that mode does not change the total energy of the state. However, a zero-energy mode in modular Hamiltonian should have a different physical interpretation. 

To gain some physical intuition, we consider the simplest situation by writing $\hat{\rho}\propto e^{-\epsilon\hat{c}^\dag\hat{c}}=\ket{0}\bra{0}+e^{-\epsilon}\ket{1}\bra{1}$. It is very clear that when $\epsilon=0$, two states, whose particle numbers differ by one, share an equivalent classical probability. Therefore, in contrast to the zero mode in physical Hamiltonian,  a zero mode in modular Hamiltonian means that adding a particle at that mode does not change the classical probability of the state in the mixed ensemble. In other words, the spectrum of the density matrix has a degeneracy. Therefore, the physical consequence of zero modes of the modular Hamiltonian is embedded in particle number statistics, and it becomes natural to consider the full counting statistics (FCS) in order to reveal nontrivial topology in mixed state.   

The generating function of FCS, also called disorder operator, is defined as \cite{fcs1,fcs2,fcs3,fcs4,fcs5,fcs6,fcs7,fcs8,fcs9,disorder1,disorder2,disorder3,YCWang21,BBChen22,YCWang22,ZHLiu23,WJiang23}
\begin{equation}
    K(\theta)=\ln\text{Tr}(\hat{\rho}e^{i\theta \hat{Q}}), \label{FCS}
\end{equation}
where $\hat{Q}$ is the total particle number operator. Here $\theta$ is a parameter restricted to $[0,2\pi)$. $K(\theta)$ is a periodic function of $\theta$ with period $2\pi$, and contains the information of all orders of cumulants of the particle number distribution function. 
We focus on the real part of $K(\theta)$, normalized by the system volume as 
\begin{equation}
    F(\theta)=\frac{1}{L^d}\Re{K(\theta)}.
\end{equation}  In this article, we will show that the FCS generating function can be used to probe nontrivial topology of a class of mixed states, namely the fermionic Gaussian states. The density matrix of fermionic Gaussian states can be generally written as
\begin{equation}
    \hat{\rho}=\frac{1}{Z}\exp{-\sum_{ij}G_{ij}\hat{c}_i^\dagger \hat{c}_j},
\end{equation}
where $\hat{c}_i^\dagger$ and $\hat{c}_i$ are fermionic creation and annihilation operators, and $Z$ is the normalization factor. We will first discuss one dimensional systems and then generalize the results to higher dimensions. 

{\it The One-Dimensional Case.} 
Before proceeding, we first summarize our main results in one-dimension:

1) If the mixed state has a nontrivial topology, $F(\theta)$ has a cusp structure
at $\theta=\pi$,
and $F^\prime(\theta)$ diverges toward $\pm \infty$ when $\theta\rightarrow \pi^{\pm}$. $F^\prime(\theta)$ is the first order derivative with respect to $\theta$.

2) This cusp structure can be translated into a topology indicator $\mathcal{I}$ defined below, whose value is quantized to unity (zero) for topologically nontrivial (trivial) mixed states. The topological indicator is defined as
\begin{equation}\label{I}
    \mathcal{I} = -\frac{1}{\pi}\int_{0}^{2\pi}\frac{d\mathcal{J}(\theta)}{d\theta} d\theta,
\end{equation}
where 
\begin{equation}\label{J}
\mathcal{J}(\theta)=\arctan(F^\prime(\theta)).
\end{equation}

3) We introduce a zero-mode counter as
\begin{equation}
    \chi(\theta)=-\left(\frac{dF(\theta)}{d\theta}\right)^2 \bigg/ \left(\frac{1}{L}\frac{d^2 F(\theta)}{d\theta^2}\right).
\end{equation}
We find that 
\begin{equation}
    \lim_{\theta\to\pi}\chi(\theta) = N,
\end{equation}
where $N$ is the number of zero modes of the modular Hamiltonian.

\begin{figure}[t!]
    \centering
    \includegraphics[width=0.48\textwidth]{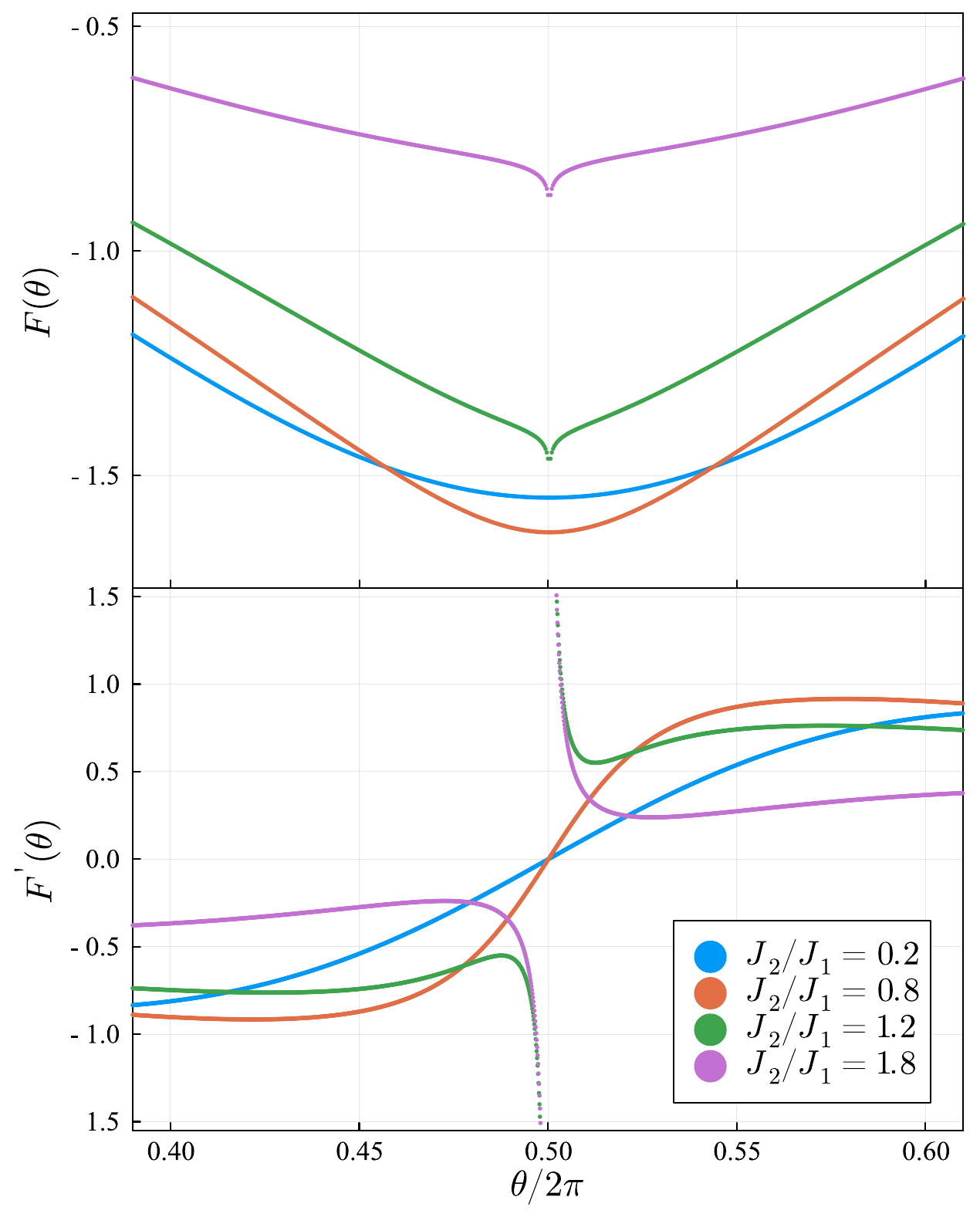}
    \caption{Upper: $F(\theta)$ for a class of one-dimensional mixed states. The modular Hamiltonians of these states are given by the form of the SSH Hamiltonian. For the topologically nontrivial mixed state ($J_2>J_1$), $F(\theta)$ exhibits a cusp at $\pi$, whereas for the topologically trivial mixed state ($J_2<J_1$), $F(\theta)$ is smooth. Lower: $F'(\theta)$ displays a divergence at $\theta=\pi$ in the topologically nontrivial mixed state; while it is continuous in the topologically trivial mixed state. We take $N=100$ unit cells for numerical simulation.}
    \label{cusp}
\end{figure}

To illustrate why these results work, we again first consider the simplest situation of a single site with $\rho=e^{-\epsilon \hat{c}^\dag\hat{c}}/Z$ and $L=1$. It is straightforward to show that 
\begin{equation}\label{F}
F(\theta)=\frac{1}{L}\Re{\left[\ln(\frac{1+e^{-\epsilon+i\theta}}{1+e^{-\epsilon}})\right]}
\end{equation}
and then 
\begin{equation}\label{F'}
    F^\prime(\theta)=-\frac{1}{L}\frac{e^{\epsilon}\sin\theta}{(e^{\epsilon}+\cos\theta)^2+\sin^2\theta}.
\end{equation}
Here, we have formally kept the factor $1/L$ with $L=1$ to remind the readers that we will properly take into account finite size effect later.
If $\epsilon \neq 0$, $F^\prime(\theta)$ is finite everywhere and is a continuous function of $\theta$. If $\epsilon=0$,  $F^\prime(\theta)$ diverges at $\theta=\pi$. And when $\theta\rightarrow \pi^{\pm}$, $F^\prime(\theta)$ approaches $\pm \infty$, respectively. 

Moreover, consider a one-dimension gapped topological chain in a finite size system with length $L$ with the presence of one zero-mode. The zero-mode energy is not exactly zero, but exponentially small in system size as $\epsilon\propto e^{-L}$. 
Let $\delta\theta=\theta-\pi$, $F'(\theta)$ behaves as follows
\begin{equation}\label{asymptotic}
    F'(\theta)\approx\left\{
    \begin{array}{cc}
    \frac1L\frac{\delta\theta}{\epsilon^2}+\text{reg},     & \mbox{$|\delta\theta|\ll|\epsilon|$}; \\
    -\frac{1}{2L}\tan{\frac{\theta}{2}}+\text{reg},     & \mbox{$|\delta\theta|\gg|\epsilon|$},
    \end{array}
    \right.
\end{equation}
where ``reg" accounts for the contributions of regular terms, i.e., the eigenmodes that are not topological zero-mode. Such terms have finite contributions since the contribution of each term is finite. Extrema are reached at $|\delta\theta|\sim |\epsilon|\sim e^{-L}$ with $|F'(\theta)|\approx 1/(L\epsilon) \sim e^{L}/L$.
Therefore, a divergence exists in the thermodynamic limit $L\to\infty$ and
$ \lim_{\theta\to \pi^\pm}F'(\theta)=\pm \infty$. 
A more detailed derivation of the cusp structure is presented in the Supplementary Material.

One can regularize this divergence by applying the inverse tangent function $\mathcal{J}(\theta)=\arctan(F'(\theta))$. For a realistic Gaussian state, all the eigenmodes of the modular Hamiltonian contribute additively to $F(\theta)$. All the finite energy modes together contribute a finite value to $F^\prime(\theta)$ in the thermodynamic limit and each mid-gap zero-energy mode contributes a divergent term behaving as Eq. (\ref{asymptotic}) at $\theta\rightarrow\pi$. If there exists a total of $N$ mid-gap zero-modes, Eq. (\ref{asymptotic}) will be multiplied by a factor of $N$.  In the absence of zero modes, $F'(\theta)$ is regular and so is $\mathcal{J}(\theta)$, and therefore, we have $\mathcal{I}=0$ due to the periodicity of $\mathcal{J}(\theta)$. However, in the presence of zero modes, $\mathcal{J}(\theta)$ jumps from $-\pi$ to $\pi$ at $\theta=\pi$, resulting in $\mathcal{I}$ quantized to unity. Therefore, a non-zero $\mathcal{I}$ can be used as an indicator for nontrivial mixed state topology. 

\begin{figure}[t!]
    \centering
    \includegraphics[width=0.48\textwidth]{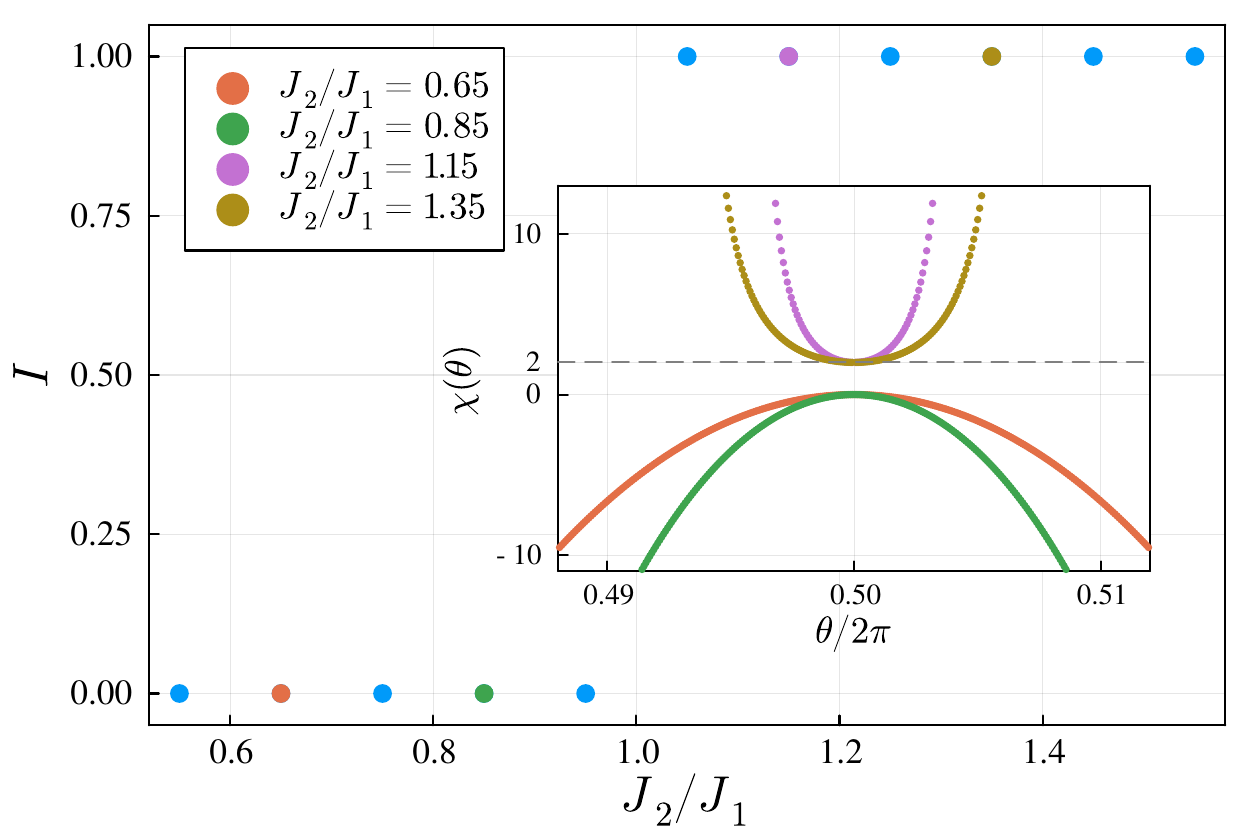}
    \caption{Topology indicator $\mathcal{I}$ for a class of one-dimensional mixed states whose modular Hamiltonians are given by the form of the SSH Hamiltonian. For topologically trivial mixed states ($J_2<J_1$), $\mathcal{I}$ is quantized to zero. For topologically nontrivial mixed state ($J_2>J_1$), $\mathcal{I}$ is quantized to unity. Inset: the zero-mode counter $\chi(\theta)$ as function of $\theta$ computed with the parameter the same as the colored dots in the main figure. $\chi(\pi)=2$ ($=0$) for $J_2>J_1$ ($J_2<J_1$) where the modular Hamiltonian has two (zero) zero modes at the edge.   We take N = 100 unit cells for numerical simulation.}
    \label{TopoInd}
\end{figure}

Note that the quantization of $\mathcal{I}$ does not depend on the number of zero modes, because it only probes the divergence in $F'(\theta)$, independent of the value of $N$. Hence it lacks the ability to identify the value of topological invariants, or the numbers of zero modes. To this end, we propose a refined quantity $\chi(\theta)$ that directly detects the number of zero modes, which we call the zero-mode counter 
\begin{equation}\label{chi}
    \chi(\theta) = -\lim_{L\to\infty}\frac{L(F'(\theta))^2}{F''(\theta)}.
\end{equation}
For a single zero mode, by using Eq. (\ref{F'}), we arrive at
\begin{equation}\label{F''}
    F''(\theta) = -\frac{1}{L} e^{\epsilon}
    \frac{(e^{2\epsilon}+1)\cos\theta+2e^{\epsilon}}{[(e^{\epsilon}+\cos\theta)^2+\sin^2\theta]^2}.
\end{equation}
In the definition of $\chi(\theta)$, the factor $L$ is eliminated, thus we can easily take the thermodynamic limit and the mid-gap states can be taken as exact zero energy states in the thermodynamic limit. For a general Gaussian state with $N$ zero-modes in its modular Hamiltonian, it is straightforward to show that  
\begin{equation}
    \chi(\theta)=\frac{N}{2}\left(\tan(\frac{\theta}{2})\right)^2(\cos\theta+1)+\dots.
\end{equation}
Here the first term comes from zero modes and ``$\dots$" represents contributions from finite energy terms, which  vanishes identically at $\theta=\pi$ because $F'(\pi)=0$ if $\epsilon\neq 0$, as shown in Eq. (\ref{F'}).
Therefore, $\chi(\theta=\pi)$ is simply given by the first term and 
$ \chi(\theta)\big|_{\theta\to\pi}=N$.

To illustrate these three results concretely, we choose a one-dimensional density matrix $\hat{\rho}=e^{-\hat{G}}/Z$ where the modular Hamiltonian $\hat{G}$ is chosen as the celebrated Su-Schrieffer-Heeger (SSH) model \cite{SSH}
\begin{equation}\label{modularH}
    \hat{G}=\sum_{i=1}^L J_1 \hat{c}^\dagger_{iB} \hat{c}_{iA}+J_2 \hat{c}^\dagger_{i+1A}\hat{c}_{iB}+\text{h.c.},
\end{equation}
where $A$, $B$ are two sublattice indices, and $\text{h.c.}$ stands for the Hermitian conjugate. The mixed state given by the above density matrix is topologically nontrivial for $J_2>J_1$ and trivial for $J_2<J_1$. For $J_2>J_1$, Fig. \ref{cusp}  shows that $F(\theta)$ has a cusp and $F^\prime(\theta)$ diverges toward $\pm\infty$ at $\theta=\pi$. Fig. \ref{TopoInd} shows $\mathcal{J}=1$ and $\chi(\pi)=2$ indicating two zero modes at two edges. In contrast, for $J_2<J_1$, Fig. \ref{cusp} shows that both $F(\theta)$ and $F^\prime(\theta)$ are smooth, and Fig. \ref{TopoInd} shows that both $\mathcal{J}$ and $\chi(\pi)$ equal zero.

Before concluding the one-dimensional case, let us also point out that the above discussions only probe zero modes and cannot determine whether the zero mode are localized at the edge. To this end, one can further modify Eq. (\ref{FCS}) by replacing $\rho$ with $\rho_\text{o}-\rho_\text{p}$ as
\begin{equation}
   K(\theta)=\ln\text{Tr}((\rho_\text{o}-\rho_\text{p})e^{i\theta \hat{Q}}), \label{FCS-edge}
\end{equation}
 where $\rho_\text{o}$ and $\rho_\text{p}$ are the same mixed state defined in open and periodic boundary conditions, respectively. This modification eliminates the contributions from the bulk modes and the modified $K(\theta)$ only counts eigenmodes of modular Hamiltonian localized at the edges.

{\it The Higher Dimensional Case.}  
When generalizing the results to higher dimensions, we need to first note on some key differences between one-dimension and higher dimensions.
In one-dimension there are a discrete number of mid-gap states in the topological phase, whose energies vanish exponentially with system size $L$ as $e^{-L}$. As we have seen in the discussion below Eq. (\ref{asymptotic}), this exponential behavior leads to a term of $O(e^L/L)$, which is crucial for the divergence in $F^\prime(\theta)$. 

In higher dimensions $d\geq2$ of size $L^d$, the gapless boundary modes have linear dispersion \cite{Bernevig}. In the following, we focus on the case that the chemical potential of the modular Hamiltonian is fixed at $\mu=0$, such that the gapless boundary states are filled up to zero energy. Subjected to the finite size effect, the zero-energy states vanish as $1/L$, and due to finite density-of-state, {the number of states of energy within $O(1/L)$ is  of  $O((\ln{L})^{d-1})$. Following similar analysis as Eq. (\ref{asymptotic}), since $\epsilon\propto 1/L$, $F^\prime(\theta)$ reaches extrema at $|\delta\theta|\sim |\epsilon|\sim L^{-1}$ with the maximum value $\sim 1/L^{d-1}$. Multiplied by the density-of-state, the contribution of these gapless surface modes to $F^\prime(\theta\rightarrow\pi)$ should be $O((\ln{L})^{d-1}/L^{d-1})$, which vanishes in the thermodynamic limit. Therefore, unlike the one-dimensional case, the gapless surface modes do not contribute a divergence in $F^\prime(\theta)$. 

A simple modification is to multiply $F(\theta)$ by $L^d$ such that this contribution becomes $O(L(\ln{L})^{d-1})$, exhibiting divergence at $\theta\rightarrow \pi$. However, this simple modification does not work because the bulk contribution to $F^\prime(\theta)$ also diverges after multiplying $L^d$. Hence, we introduce $g(\theta)$ function as  
\begin{equation}
    g(\theta)=[F^\prime_\text{o}(\theta)-F^\prime_\text{p}(\theta)]L^d, \label{gtheta}
\end{equation}
where $F^\prime_\text{o}(\theta)$ and $F^\prime_\text{p}(\theta)$ respectively denote $F^\prime(\theta)$ defined with $\rho_\text{o}$ and $\rho_\text{p}$, respectively. As discussed above, this definition eliminates the bulk contributions and displays a jump of $O(L(\ln{L})^{d-1})$ at $\theta=\pi$. Consequently, we can define the topological indicator similar as Eq. (\ref{I}) and (\ref{J}) with $F^\prime(\theta)$ replaced by $g(\theta)$.  

\begin{figure}[t!]
    \centering
    \includegraphics[width=0.48\textwidth]{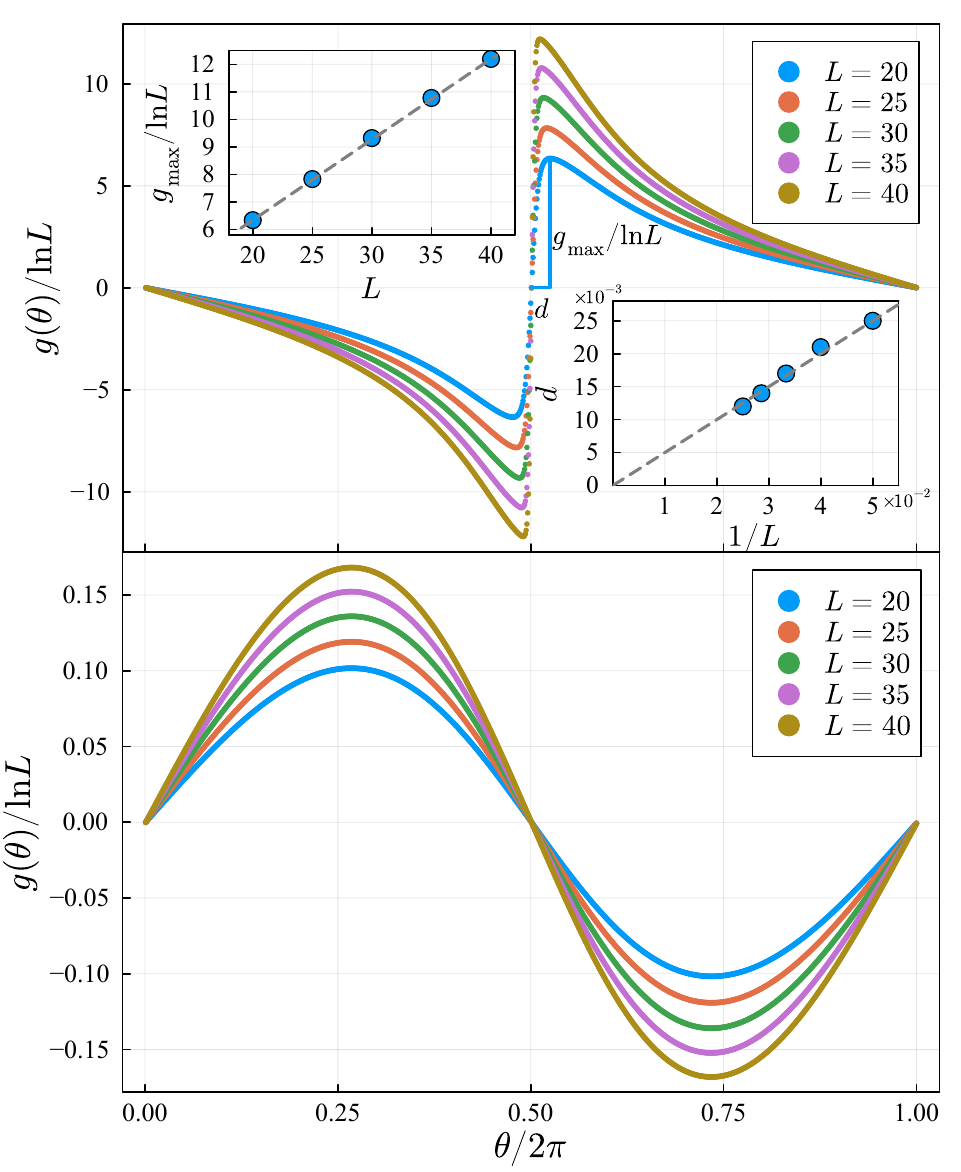}
    \caption{The behavior of $g(\theta)$ defined by Eq. (\ref{gtheta}) for two-dimensional mixed states. $g(\theta)/\ln L$ as a function of $\theta$ is plotted for different linear system size $L$. The modular Hamiltonian is taken as the Haldane model. Upper panel: Topological phase with $\mu/J_2=0.0$ and $\phi=\pi/2$. Inset: The height of the jump $g_\mathrm{max}/\ln L$ as function of linear system size $L$, and the width of jump $d$ as function of $1/L$. Lower panel: Trivial phase with $\mu/J_2=8.0$ and $\phi=\pi/2$. No cusps form near $\theta=\pi$}
    \label{2d}
\end{figure}

We numerically verify the above analysis in two-dimension by calculating a Gaussian mixed state whose the modular Hamiltonian is given by the Haldane model on honeycomb lattice \cite{Haldane88}
\begin{align}
    \hat{G}&=J_1\sum_{\langle ij\rangle}\hat{c}^\dagger_i \hat{c}_j
    +h.c.+J_2\sum_{\langle\langle ij\rangle\rangle}
    e^{\pm i\phi}\hat{c}^\dagger_i \hat{c}_j+h.c.\notag\\
    &+\mu(\sum_{i\in A}\hat{c}^\dagger_i \hat{c}_i-\sum_{i\in B}
    \hat{c}^\dagger_i \hat{c}_i),
\end{align}
where $\langle \rangle$ and $\langle\langle\rangle\rangle$ denote summation over the nearest and the next nearest pairs of sites, respectively. The phase $\phi$ breaks time-reversal symmetry and takes plus (minus) sign for anti-clockwise (clockwise) next nearest hopping. We chose $J_1$, $J_2>0$. 
When $\mu/J_2<3\sqrt{3}|\sin\phi|$, the model is in the topological phase with Chern number $C=+1$ for $0<\phi<\pi$ and $C=-1$ for $-\pi<\phi<0$. We numerically compute the behavior of $g(\theta)$ for the topological phase $\phi=\pi/2$ and $\mu/J_2=0$, and the numeric results are shown in Fig. \ref{2d}. Here the open boundary means stripe geometry and the periodic boundary means torus geometry, where the length of the stripe or the torus is denoted by $L$. It shows that $g(\theta)$ is anti-symmetric around $\theta=\pi$ and displays a discontinuity at $\theta=\pi$. The inset shows the jump scales with $L\ln L$ and the width of the jump scales with $1/L$. This shows that $g(\theta)$ defined in Eq. (\ref{gtheta}) for higher dimension shares the same behavior as $F^\prime(\theta)$ defined for one-dimension. 

{\it Remarks and Outlook.} 
In summary, the density matrix of a mixed state can be equivalently described by its modular Hamiltonian. When the modular Hamiltonian has a nontrivial bulk topology, the stable zero energy edge states guaranteed by the bulk-boundary correspondence lead to a degeneracy in the density matrix spectrum. This work shows that this degeneracy has dramatic physical consequences in the full counting statistics. Remarkably, The full counting statistics can be experimentally measured. For instance, for ultracold atoms in the optical lattice, using the quantum gas microscope one can directly measure the occupation of neutral atoms up to single atom precision \cite{fcs_exp1,fcs_exp2,fcs_exp3}. This measurement basis is exactly the diagonal basis of $e^{i\theta\hat{Q}}$. By repeating such measurements multiple times, one can get the ensemble average of $e^{i\theta \hat{Q}}$, and then obtain $F(\theta)$.

Our current discussion is restricted to Gaussian mixed states, it is interesting to consider the case where the modular Hamiltonian is of the interacting form and the mixed state is no longer Gaussian. The effect of interactions on full counting statistics remains an open question and deserves more study in the future.
Furthermore, it is also conceivable that full counting statistics may be used to detect mixed states that host nontrivial topological orders \cite{Yin23,Bao23,Fan23,aspt1,aspt2,cenke1,cenke2,interaction1,interaction2,interaction3,interaction4,interaction5}. 

\begin{figure}
    \centering
    \includegraphics[width=\columnwidth]{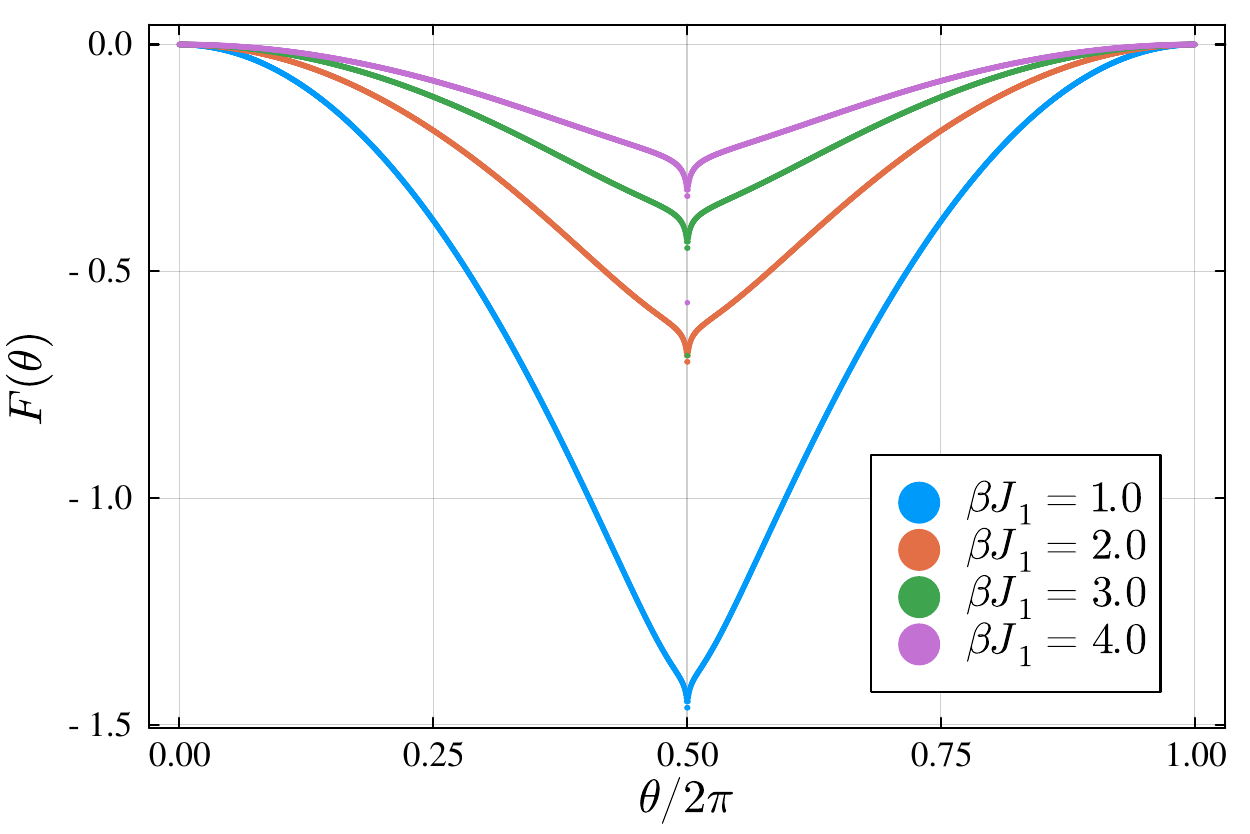}
    \caption{The singluarity peak in $F(\theta)$ remains robust for a wide range of temperatures.}
    \label{temp}
\end{figure}

In principle, $F'(\theta)$ and $g(\theta)$ also contain singularities when accidental zero-energy modes are present in a topologically trivial state. However, these zero-energy modes are susceptible to local disorders and therefore typically not present.
We also note that our results can be applied to topological band insulators at finite temperatures. Our results state a highly nontrivial fact that quantization can be observed for a topologically nontrivial Hamiltonian at any finite temperature.
As illustrated in Fig. \ref{temp}, the cusp persists for a wide range of temperatures showing a remarkable stability against thermal effects. In conventional wisdom, the physical observables that characterize the ground state topology is no longer quantized at finite temperature. However, $F(\theta)$ is a fundamentally different type of observable that is a nonlinear function of $\hat{\rho}$ making it insensitive to temperature.

As mentioned in the introduction, quantized observables for specific symmetry classes have been proposed for some one- and two-dimensional fermionic systems at finite temperatures.
For one-dimensional Gaussian fermionic systems with chiral symmetry, the ensemble geometric phase \cite{top_mod1} was proposed to detect the mixed state topology.
For two dimensions, generalization of Chern number and $\mathbb{Z}_2$ topological invariant for mixed states also exist \cite{top_mod2,top_mod4}. These topological invariants exactly detect the band topology of modular Hamiltonians. However, there only exist experimental protocols for measuring ensemble geometric phase. How to measure the topological invariant for 2D mixed states remains to be elucidated. Our proposals, on the other hand, detect the topological edge states on the boundary.
  The singularity in FCS can be measured in both 1D and higher dimensions. Although this singularity cannot tell us the value of the topological invariant, it serves as a diagnosis of whether the density matrix is topologically nontrivial. Furthermore, in 1D the zero-mode counter gives the value of the topological invariant through bulk-boundary correspondence.
The relation between those observables and the quantization in FCS at finite temperatures remains unclear and needs to be further explored.  

Open source code is available at \cite{code}. The method is exact diagonalization.

\textit{Note Added.} When preparing this manuscript, we became aware of several related works where the cusp structure of $F(\theta)$ at $\theta=\pi$ was also reported for different physical contexts \cite{yingfei, pengfei, meng}.

{\it Acknowledgments}
     We thank Yingfei Gu, Pengfei Zhang, Zhen Bi, Chong Wang, Chengshu Li, and Xingyu Li for helpful discussions.  This work is supported by the National Key R\&D Program of China 2023YFA1406702, the Innovation Program for Quantum Science and Technology 2021ZD0302005, and the XPLORER Prize. This work is also partly supported by a new faculty member starting grand at Southeast University No.4007022417.

\end{document}